\documentclass[11pt,titlepage]{article}

\usepackage{graphicx}
\usepackage{color}

\usepackage{setspace} 
\usepackage{epsfig}

\usepackage{amsmath}

\usepackage{graphicx}

\definecolor{red}{rgb}{1,0,0}
\newcommand {\hide}[1]{\null}
\newcommand {\D}{{\rm d}}
\newcommand {\kb}{k_{\rm B}}

\newcommand {\grad}{\nabla}
\long\def\symbolfootnote[#1]#2{\begingroup%
\def\thefootnote{\fnsymbol{footnote}}\footnote[#1]{#2}\endgroup}
\def\blfootnote{\xdef\@thefnmark{}\@footnotetext}
\renewcommand{\thefootnote}{\fnsymbol{footnote}}

\title{{New Proposed Mechanism of Actin-Polymerization-Driven Motility}}

\author{Kun-Chun Lee\\ 
	Department of Physics and Astronomy, \\
	University of Pennsylvania, Philadelphia, PA.
      \and Andrea J.~Liu 
         \thanks{
         Corresponding author.  Address: 
	   Department of Physics and Astronomy,
         University of Pennsylvania,
	   205 S. 33rd Street,
	   Philadelphia, PA. 19104--6396, U.S.A.,
	   Tel.:~(215)573-7374, Fax:~(215)898-2010} \\
	Department of Physics and Astronomy, \\
	University of Pennsylvania, Philadelphia, PA.}

\date{}

\begin{document}

\maketitle

\abstract{

We present the first numerical simulation of actin-driven propulsion by elastic filaments.  Specifically, 
we use a Brownian dynamics formulation of the dendritic nucleation model of actin-driven propulsion.  
We show that the model leads to a self-assembled network that exerts forces on a disk and pushes 
it with an average speed.  This simulation approach is the first to observe a speed that varies non-monotonically with 
the concentration of branching proteins (Arp2/3), capping protein and depolymerization rate (ADF), in accord with experimental observations.  
Our results suggest a new interpretation of the origin of motility that can be tested readily by experiment.  

}

\clearpage

\section*{Introduction}
There is a type of biological motility, used in a form of cell crawling (1) and by intracellular pathogens such as {\it Listeria monocytogenes}(2), that is driven not by motor proteins but by biological self-assembly of the protein actin.  During this process, ATP hydrolysis and activation of the protein complex Arp2/3 drive actin self-assembly from monomers (G-actin) to branched networks of filaments (F-actin)  (3), thus providing the necessary thermodynamic
free energy to push a bacterium or a cell forward (4,5).  This driven, non-equilibrium self-assembly process is regulated by a cadre of proteins.  It is now possible to drive a latex bead through a buffer solution containing only these proteins (6-9).  Such beads travel through solution propelled by a dense branched actin network at their rear, demonstrating that non-equilibrium self-assembly of F-actin is sufficient to drive motility.

The standard biochemical model 
for the regulation of actin-self-assembly-driven motility
is the dendritic nucleation model (3,10,11).  In this model, actin self-assembles (or "polymerizes") into filaments preferentially at one end (the barbed end) and "depolymerizes" preferentially at the other end (the pointed end).
Proteins such as WASP at the moving surface (the rear end of the Listeria bacterium or moving latex bead, or the leading edge of the membrane of a crawling cell)
recruit and activate the Arp2/3 protein complex.
The activated Arp2/3 catalyzes the nucleation of new branches from pre-existing actin filaments, thus creating new growing barbed ends near the moving surface.
To sustain motion, two other essential proteins regulate the turnover of actin monomers:
severing protein (ADF), which raises the depolymerization rate by severing filaments in two, and capping protein (Cap), which covers barbed ends and prevents further 
growth.  Thus, filaments just behind the moving surface at the front of the branched network tend to grow due to Arp2/3 and
WASP, while filaments at the far end of the branched network tend to depolymerize away
due to ADF and Cap.  

A key physical question that arises is:  how does the self-assembly of a branched network generate forces and produce motion?  Many models have been developed to show how the polymerization of a single actin filament can produce a force (12-22).   Other models show how the dendritic nucleation model creates a branched network morphology (23-28).  Relatively few models have considered how polymerization of a branched network might lead to force generation; of these, some treat the network as an elastic continuum (29-31).  Only three approaches explicitly incorporate the morphology of the dendritic nucleation model to produce force and motion (28,32,33).
In all three of these
simulation models, mass is not conserved; monomers spring into existence and become capable of exerting forces only when they join filaments, and vanish when they fall off.   As a result, matter is created just behind the moving surface, leading to motility as an artifact.

In this paper, we use Brownian dynamics to demonstrate that force and motion can indeed emerge from the growth of a branched network in a physically-consistent model.  We demonstrate that our model is the first to capture key properties of the dendritic nucleation model by reproducing the characteristic dependence of speed on the
concentrations of Arp2/3, capping protein, ADF and actin (8,34,35).  Our simulation suggests a new understanding of the mechanism driving motility: the disk emits activated Arp2/3 complex, which gives rise to a buildup of F-actin just behind the disk.  If there is a repulsive interaction between the disk and the actin, the disk will move forward to avoid the actin recruited by Arp2/3.  We propose explicit experiments to test this new picture.

\section*{ Simulation Model}

Relevant time scales within the dendritic nucleation model
span six orders of magnitude.  The longest time scale is set by kinetic events
such as the depolymerization rate ($\sim 1$s) (3), while the shortest time scale is determined by
diffusion and collision of G-actin monomers ($\sim 1\mu$s) and the high frequency dynamics of 
filaments.  The wide range of important time scales poses a challenge to computer simulation.  Previous approaches avoid this problem by treating free monomers and those in filaments very differently, leading to potential artifacts (36).    If one insists on treating free monomers and monomers within filaments consistently, one must use a time step that is small enough to capture their short-time-scale dynamics when integrating the equations of motion.  On the other hand, in order to study the steady-state, one must be able to reach time scales that are long compared to the slowest reaction rate involved.  The compromise that we have chosen is to narrow the range of time scales by increasing the slowest rates, such as the depolymerization rate, by 5 orders of magnitude and and decreasing the filament stiffness by one order of magnitude (see Table 1).   We also use the enhanced depolymerization rate to mimic the action of severing protein (ADF).    The details of our model are presented in the Methods section.

To offset some of these changes, we adjust other variables so that the steady state fluxes are comparable to those observed experimentally.
For example, to offset the effect of our artificially-high depolymerization rate, 
we increase the typical concentration of the G-actin monomers such 
that the ratio of the effective polymerization rate 
and the depolymerization rate, $K_{+}$[G-actin]$/K_{-}$, is close
to the typical experimental value.

In testing our model, our aim is not to reproduce numerically accurate results but to capture experimentally-observed trends and understand what factors control them.  In particular, our goal is to gain insight into the mechanism that leads to motility.  We will show that the origin of motility in our system
(see Results) suggests a possible mechanism for the real system that 
yields a reasonable speed within a simple order-of-magnitude estimate 
(see Discussion).

In order to check that our conclusions do not result from the unphysical parameters we have chosen, we have varied the parameters over a range.  For example, most of our runs were carried out for a bending stiffness of $K_B=100\kb T$ (see Eq.~\ref{stretch}), corresponding to a persistence length of 0.5 $\mu$m.  However, we have also shown that when all other parameters are held fixed, we obtain the same speed for $K_B=1000\kb T$, corresponding to a more realistic persistence length of $5 \mu$m.

We have also checked the dependence of our results on $K_-$ and other slow rates by decreasing them and showing that the trends remain the same.

\section*{Bulk system}  Our simulation model is described in the Methods section.  For systems that are spatially isotropic on average, we have shown that the Brownian dynamics results for morphology are in quantitative agreement with a mean-field formulation of the dendritic nucleation model (27).  This mean-field formulation was in turn shown to be in quantitative agreement with {\it in vitro} experiments (37).  Thus, our model yields reasonable results for the steady-state bulk system.

\section*{Motility}
\begin{figure}[tbh]
\centering{
\psfig{file=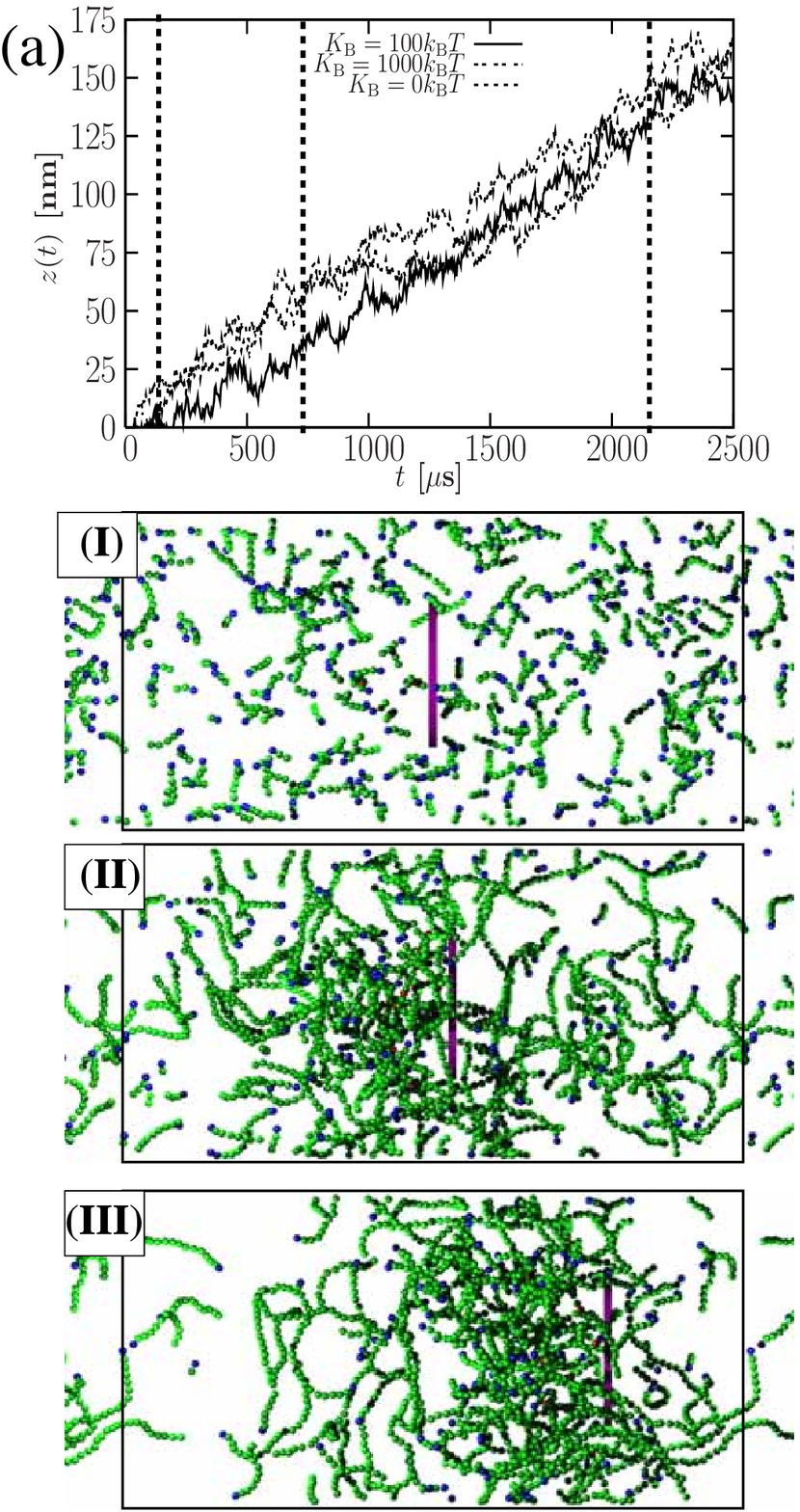,width=8.0cm,height=13cm}
}
\caption{
}
\label{fig2}
\end{figure}

We now break symmetry by introducing a moving surface in the form of a disk, whose back surface (facing the $-z$ direction) emits Arp2/3.  This drives self-assembly of a branched network behind the disk, which pushes the disk in the $+z$ direction.  We typically begin each run with 5-10\% of the actin monomers in dimer form and the rest as free monomers. We begin with some dimers as protofilaments because spontaneous nucleation of filaments, which occurs at a very low rate experimentally (3), is not allowed in our model.   We find that the results are not sensitive to the fraction of initial dimers.
The dimers and free monomers are initially distributed randomly in the system.
Fig.~\ref{fig2}(a) shows the displacement, $z$, of the disk as a function
of time for a typical simulation run for a filament stiffness $K_b=100$ (solid curve).  The dashed vertical lines mark the times corresponding to the snapshots (I-III).  In the snapshots, free monomers are not shown.  The black box corresponds to the simulation box; we have shown part of the periodic images to the right and left.  Snapshot I displays the system at 
$t = 70\mu$s.  At this time, the disk is still very close to its starting position. The dimers have grown into short filaments and are dispersed throughout the box.
By $t = 700\mu$ms (snapshot II),  a branched F-actin network has formed behind the disk and the disk has moved slightly.  By $t = 2100$ms (snapshot III), the disk has moved to the right by nearly a 
third of the simulation box.

Fig.~\ref{fig2}(a) shows that once the disk starts moving, the trajectory is linear.   Because there are significant fluctuations in the displacement (9,38), we extract speeds from trajectories that are at least $7000\mu$s long (several times longer than that shown in Fig.~\ref{fig2}), and average over the final 3500-4200$\mu$s of the trajectory (it takes roughly 1000$\mu$s to reach steady state).  The error bars for the speed in all of our figures were obtained from the standard deviation calculated over 5 separate simulations run under standard conditions (see Table 1).

The typical speed for
our simulated systems is 60$\mu$m/s.   
Our speed is simply determined by the polymerization rate.   We use $K_+=63 \mu {\rm M}^{-1}{\rm s}^{-1}$ 
(see Table 1) .  To convert this to 
a net polymerization speed $v_p$, we must 
multiply by the monomer size, $\sigma=5$nm, by the free monomer concentration just behind the disk, and a factor that characterizes the structure of the network.
A reasonable approximation to this factor is $\cos \theta$, where $\theta$ is the angle between the average tangent vector of filaments 
just behind the disk and the normal to the disk (13).  
For the conditions corresponding to Fig.~\ref{fig2}, 
[G-actin]$\approx$1 mM and $\cos(\theta) = 0.1-0.2$.  This yields an estimate of 
the polymerization speed $v_p \approx 31-63 \mu$m/s, in good agreement 
with our result.

The speed found experimentally is significantly slower, with a typical value of a fraction of a micron per minute (7-9).  We find that when we decrease the depolymerization rate and G-actin concentration by a factor of 10, leaving the ratio $K_+{\rm [G-actin]}/K_-$ fixed, the speed decreases by a factor of $\sim 10$.  As Table I shows, the value of $K_+{\rm [G-actin]}/K_-$ that we use is close to the experimental value, but $K_-$ and [G-actin] are much higher in our simulation.  We would therefore expect our speed to be too high.

It is also possible that part of the difference between our simulated speed and the experimental speed may be due to our neglect of filament binding to the moving surface.
Experimentally, it is known that filaments 
in the branched network bind to the proteins on the disk that active Arp2/3 complex (14, 29, 38-41).   Finite element simulations (42) suggest that the inclusion of a binding energy between filaments and the disk slows down the speed significantly and enhances fluctuations around the average speed.

An important and surprising result of our calculation is that the speed is independent of bending stiffness.  This is shown in Fig.~\ref{fig2}(a), where the speed is the same for systems with bending stiffnesses of $K_B=1000\kb T$, $K_B=100\kb T$ and  $K_B=0\kb T$.    Note that in the flexible case, we have not shown the initial start-up of the disk, which is substantially longer than for stiffer filaments.   Our results show that flexible and stiff filament networks exert comparable forces as the filaments polymerize. In all cases, the speed is simply the polymerization speed $v_p$.
Thus, the physical origin of motility does not depend sensitively on the bending
stiffness of the filaments, as is commonly believed (see Discussion).

\section*{Origin of Steady State Force}

\begin{figure}[tbh]
\centering{
\psfig{file=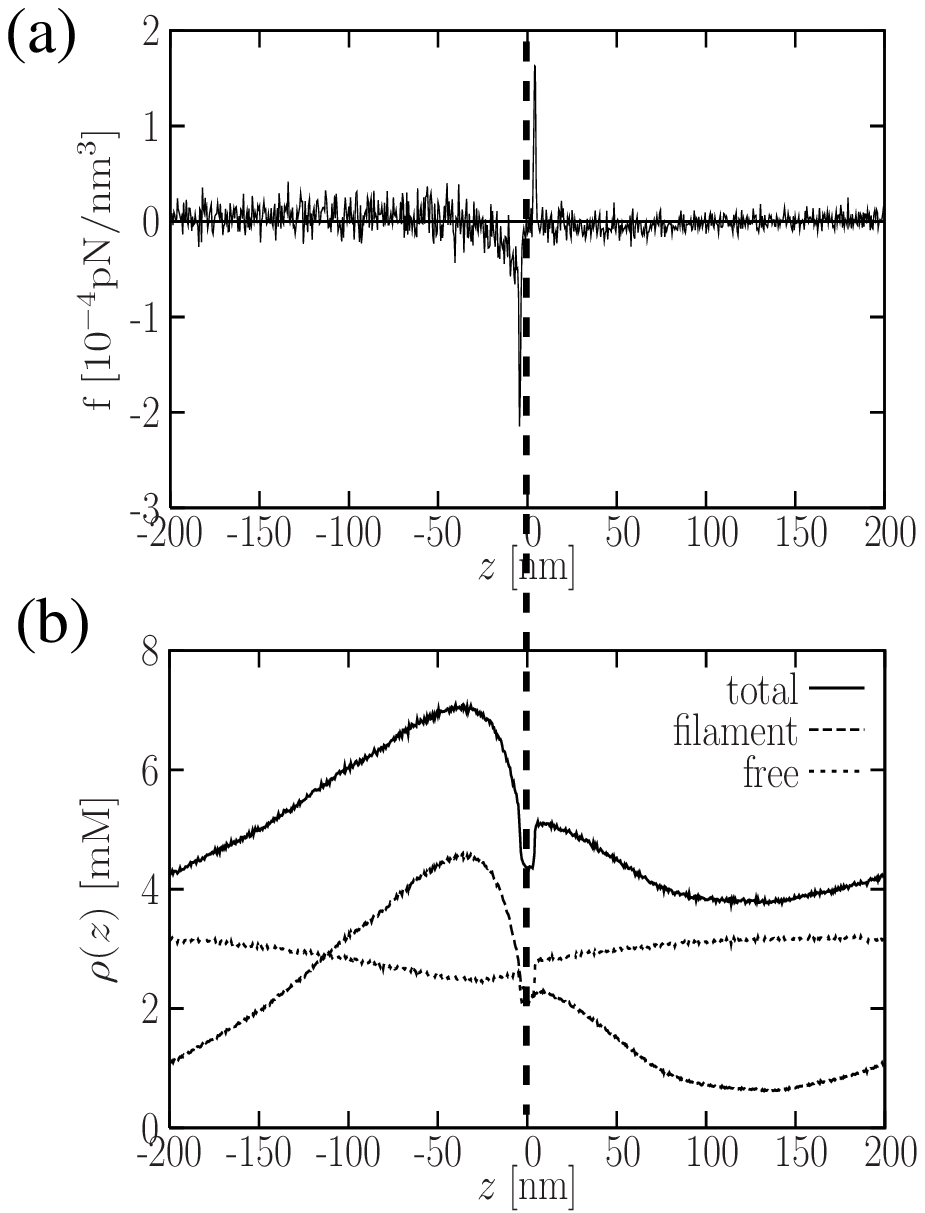,width=7.3cm,height=7.0cm}
}
\caption{
}
\label{stress}
\end{figure}

When the system is in steady state and there is a net force on the disk moving it forwards, 
there must be an equal and opposite net force on the actin.  
We have calculated the average force density $F_z(z)$  in the $z$-direction at
different distances $z$ from the disk (recall that the disk is constrained 
to move only in the z-direction).  Fig.~\ref{stress} shows 
$F_z(z)$ in the frame of the moving disk, where $z=0$ (marked by a vertical dashed line) always marks the 
position of the disk.
Just behind the disk at $z<0$, the force on monomers (free or bound 
in filaments) is large and negative, as expected because the 
force exerted by these monomers on the disk is positive.  Note that this negative force persists out to about 
50nm behind the disk before it drops nearly to zero.  For $z>0$, the 
force near the disk is positive because monomers immediately 
in front are pushed along by the disk.  We have verified 
that the total average force, $\int F_z(z)\D V$, exerted on the actin is 
equal and opposite to the force on the disk, as it must be.  
The force on the bead is 
on the order of 0.1 pN.  Although this force seems small, we note that it is the magnitude of the force required to push a $1\mu$m bead at the experimentally-observed speed, and is also, by construction, the force needed to push the disk 
at the speed that we observe for the viscosity chosen; we have confirmed that 
the average force on the disk is related to its average speed by 
the drag on the disk, $\zeta_D$, as expected.   

Note that while the negative force extends to 50nm behind the disk, the total length of the actin comet tail in our simulations is about 150nm (Fig.~\ref{fig2}.III).   Thus, only a relatively small fraction of the network directly behind the disk is subjected to a significant backwards force.  This result 
is consistent with the experimental finding that the actin network in the tail is stationary (7, 43).

We remark that the force profile shown in Fig.~\ref{stress}(a) does not contradict the experimental observation that the shape of the tail can be deformed at distances far greater than 50nm from the surface (31), because the moving surface was curved in the experiment and the tail expanded as it moved backwards away from the surface, due to entropy or elastic stresses.

The force on the disk can be viewed as the Newton's third 
law reaction force to the force in Fig.~\ref{stress}(a) on the actin, 
evaluated at the surface of the disk.  Thus, uncovering the origin of the force profile 
behind the disk should help us to understand motility.   The solid curve in Fig.~\ref{stress}(b) shows the 
density profile $\rho(z)$ of actin 
(note that the free monomer density is nearly constant, with a small dip just behind the disk, so that most of the variation is due to monomers in filament form).  
In equilibrium, similar density profiles can arise from 
attraction to the surface.  In that case, the chemical potential must 
be the same everywhere.  However, in this steady-state driven system, 
the density profile does not arise from attractions--the 
interaction of actin with the disk is purely repulsive.  
Rather, the density profile is a {\it non-equilibrium} effect, 
arising from the action of Arp2/3, which is emitted from the disk.  
(In the real system, Arp2/3 is activated at the surface of the disk, 
so the disk serves as a source of activated Arp2/3.)  
The non-equilibrium density profile leads to a pressure gradient, 
$dp/dz=(dp/d\rho)( d\rho/dz) = -(1/\kappa \rho) d\rho/dz$, 
where $\kappa$ is the local compressibility of the 
branched network.  The  importance of the compression modulus has 
been emphasized in previous models (29-31).  
In our case, the force generated depends not only on $\kappa$ 
but on the concentration gradient, $d\rho/dz$.  
Note that the pressure gradient is equal and opposite to the 
force per unit volume on the actin, 
shown in Fig.~\ref{stress}(a).  The vanishing of the force near $z=-30$nm 
therefore corresponds to the maximum in the concentration there, 
where $d\rho/dz=0$  (Fig.~\ref{stress})(b)).

We emphasize that this is not a simple osmotic pressure effect due to free monomers.  The density of free monomers (dotted curve in Fig.~\ref{stress}(b)) is nearly constant, so that the density gradient arises from F-actin, not G-actin.  

The fact that the speed corresponds to the polymerization speed for different filament stiffness suggests that the system adjusts the force exerted on the disk to maintain the speed at the polymerization speed, at least at the small loads studied here.  Therefore, it should be possible to understand the underlying mechanism for motility without involving the force.  The following interpretation does not invoke forces explicitly, but is equivalent to the above arguments and much simpler.  In this picture,
Arp2/3 complex recruits F-actin to the vicinity of the disk.  The interaction between actin and the disk is repulsive, so the disk moves forward to lower the concentration of actin near the surface.  This leads to the steady-state density profile of Fig.~\ref{stress}(b) as well as steady-state motion of the disk at the net polymerization speed.

\section*{Dependence on protein concentrations}

One key observation of experiments is that the speed is a non-monotonic function of the concentrations of the regulatory proteins involved in the dendritic nucleation model, namely Arp2/3, capping protein and severing protein.  Our simulation model is the first to capture this behavior and to explain the physical origin of the non-monotonicity.

\begin{figure}[tbh]
\centering{
\psfig{file=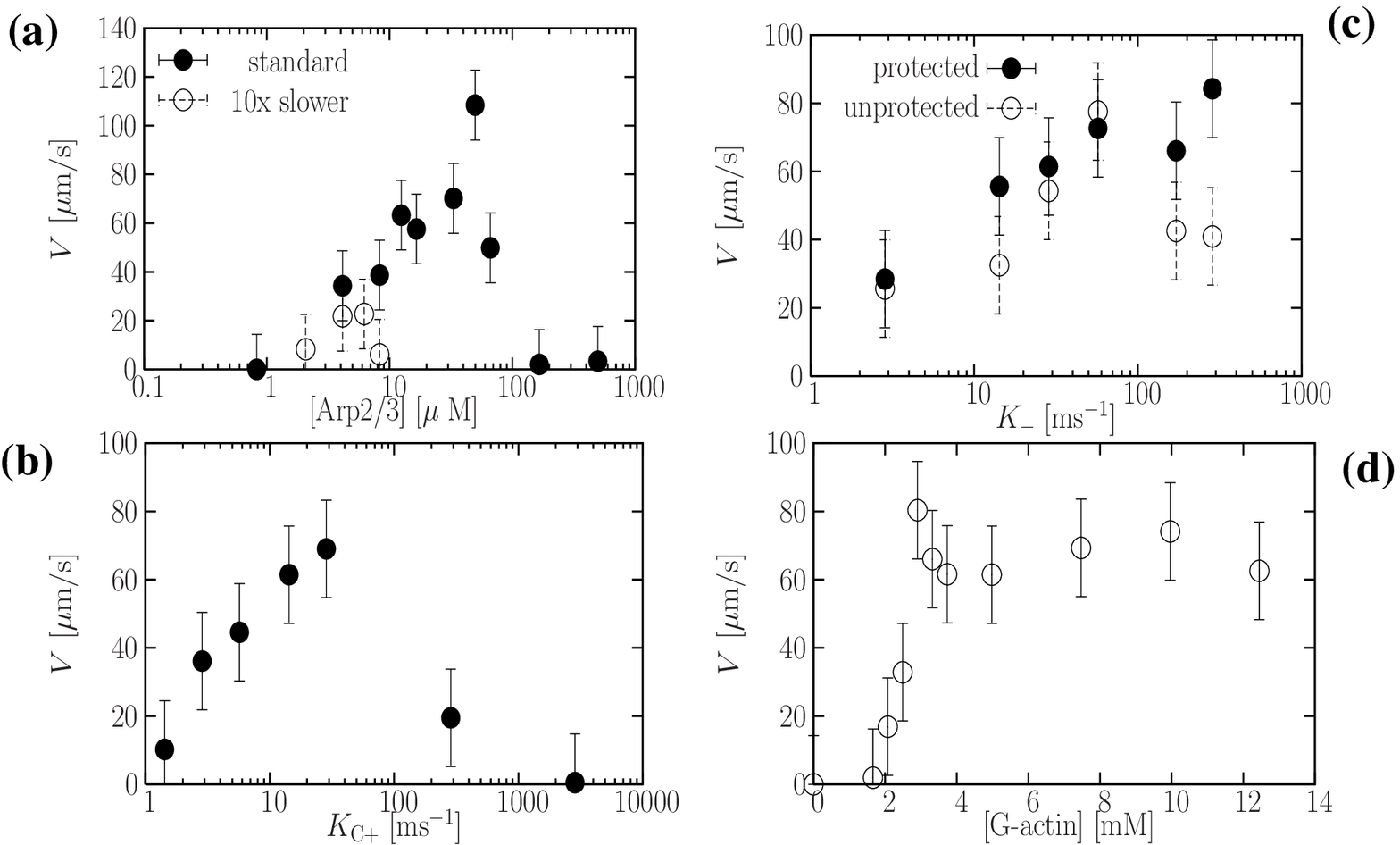,width=8.5cm,height=5cm}
}
\caption{
}
\label{dep}
\end{figure}

Fig.~\ref{dep}(a) shows that the speed is a non-monotonic function of 
Arp2/3 concentration.  Similar non-monotonic behavior has been found experimentally (8).   The behavior can be understood as follows.  At high Arp2/3 concentrations, most of the excess Arp2/3 is
trapped in filaments in the network, forming stubby branches.  These short, stubby branches do little to increase the actin concentration behind the disk.  However, they do repel actin monomers, lowering the concentration of free monomers at the surface so that fewer of them are available for polymerization.  This crowding effect is captured for the first time in our simulation because we treat monomers explicitly.  
At high Arp2/3 concentration, there appear to be two effects that reduce the speed:  first, the maximum in the density profile in Fig.~\ref{stress}(b) broadens as stubby branches proliferate.  Second, the concentration of G-actin at the surface decreases.  With increasing Arp2/3 concentration, the G-actin concentration at the surface drops below its critical value for polymerization and/or the density gradient in F-actin vanishes; at this point, the speed drops to zero.

The open symbols in Fig.~\ref{dep}(a) show the speed as a function of [Arp2/3] in the case where the depolymerization rate, debranching rate and G-actin concentration have all been decreased by a factor of 10 and the capping rate has been decreased by a factor of 5, relative to the values in Table 1.   While still high, the difference between the closed and open symbols shows the trend to be expected if we could reduce the parameters to their experimental values.  The overall trends are the same in both cases, but the speed is slower, as discussed earlier, and the maximum speed is at lower Arp2/3 concentration, as one might expect.   The maximum is much narrower as a function of [Arp2/3], which is more consistent with experimental results (8).

We note that for the typical parameters listed in Table 1 as well as the reduced parameters,  the net polymerization rate is comparable to that in the real system.  As a result, the filament density in the comet tail is comparable to that observed experimentally.  We find a filament density of order a few mM in the comet tail for our standard runs, and of order 0.35 mM for the runs with the reduced reaction rates.  This shows that there is some decrease in the amount of F-actin in the comet tail with decreasing depolymerization rate, but the values we find compare reasonably well with previous results of Carlsson (28).  One can also estimate the filament density from the Young's modulus, measured to be $Y=10^{3}$ Pa (29).  The Young's modulus for a network of semiflexible polymers with persistence length $\ell_p$ and mesh size $\xi_m$ is (44, 45)
\begin{equation}
Y=k_BT \ell_p^2/\xi_m^5
\end{equation}  
where $\xi_m=1/\sqrt{\sigma c}$, where $\sigma$ is the filament diameter and $c$ is the monomer concentration.  This yields $c \approx 1$mM, as well.

Since the filament density in our simulation is approximately the same as that in experiments, it is reasonable that the monomer concentration should be reduced near the surface relative to its value in the bulk in the real system.  This reduction inevitably leads to a reduction of the polymerization rate with increasing Arp2/3 concentration.  

Fig.~\ref{dep}(b) shows that the speed is also non-monotonic with capping protein, in agreement with experiment (8, 34).  In this case, it is obvious that too much capping will lead to a vanishing speed.  If the capping rate is too low, however, the speed also vanishes.  This is because capping and branching act synergistically.  Capping stops free barbed
ends from growing, thus forcing the system to favor branching to generate new growing ends instead of merely lengthening existing filaments (46, 34, 27).   The capping rate at the maximum of the curve in Fig.~\ref{dep}(b) is comparable to the debranching rate.

Fig.~\ref{dep}(c) shows the dependence of speed on 
the depolymerization rate.  The closed circles correspond to the case in which  the Arp2/3 protects the pointed end from depolymerization once it reaches a branch point, and prevents the branch from 
falling off.  There is experimental evidence that Arp2/3 protects 
the pointed end from depolymerization (10, 47).   
In this case, Fig.~\ref{dep}(c) shows that the speed saturates with increasing depolymerization rate.  
The open circles correspond to the case in which depolymerization proceeds through the branch point, and the branch 
falls off.   This is consistent with experiments that show that when 
ADF cofilin is present, Arp2/3 no longer protects the pointed end from depolymerization
(47).  Fig.~\ref{dep}(c) shows 
that in this case, the speed is non-monotonic and decreases with 
sufficiently high depolymerization rate.   
The experiments of Loisel et al. (8) 
exhibit non-monotonic dependence, similar to the open circles in 
Fig.~\ref{dep}(c).  This suggests that ADF does indeed prevent Arp2/3 
from protecting the pointed end from depolymerization.  Note that the two 
curves are the same at low $K_-$, and begin to deviate from 
each other near the maximum.  This corresponds to where $K_-$ 
is comparable to the debranching rate.  

Finally, Fig.~\ref{dep}(d) shows the dependence on the overall
actin concentration. Again, our results are qualitatively consistent 
with those of 
experiment (35).
The speed increases with [Actin], because the polymerization 
rate increases, and saturates at high [Actin] at a maximum polymerization speed, $v_p$.  At high [Actin], the concentration of free actin monomers at the surface, needed for polymerization, saturates at roughly 1 mM.  This saturation apparently occurs because the branched network becomes denser and more difficult for the free monomers to penetrate in order to reach the disk (24,48, 49).

\section*{Discussion}
Our simulations show that a physically reasonable formulation of the dendritic nucleation model can lead to motility.  We have taken great care to avoid possible artifacts.  For example, we treat free monomers at the same level as monomers in filaments so that there is no artificial mass transfer or dynamical discontinuity when monomers join or leave filaments.   

We have shown that the speed does not depend on the bending stiffness of the filaments (33, 50).  This surprising result appears to be consistent with the observation that amoeboid sperm of nematodes (50, 51) moves using  
a structurally different filament composed 
of major sperm protein (MSP) instead of actin.  These MSP filaments assemble into thick bundles (52) which are likely to be much stiffer than actin filaments. 

The case we have studied should correspond to the elastic ratchet model without attached filaments (13) because we have not included binding of filaments to the disk.  
However, our results appear to be at odds with the elastic ratchet model, which should predict a speed that depends on filament stiffness.  One possible source of the discrepancy is that the elastic ratchet model assumes that all of the force on the disk is exerted by monomers at the barbed end of filaments.  In our simulations, roughly 40\% of the force applied to the disk by filaments arises from monomers that are not at the barbed ends for the stiffer filaments we have studied. 

Why then is the speed insensitive to the bending stiffness of filaments?
Recall that the disk repels actin, so it prefers a low concentration of actin near the surface.  It keeps the actin concentration near its surface low by constantly moving forwards, away from the build-up of F-actin due to the action of Arp2/3.  In this picture, the bending stiffness of filaments is not particularly important to the speed, at least at small loads.   However, it is likely that the bending stiffness is important to other attributes of motility, such as the ability to withstand high loads.

This new way of thinking about the origin of motility suggests that other experimental realizations of motility should be possible.  Any system that can create a non-equilibrium, steady-state concentration profile should be able to develop a steady-state speed.  In a real system, the mechanism is somewhat different because of fluid flow (53).  The non-equilibrium chemical potential gradient resulting from the concentration gradient will lead to fluid flow, which will in turn push the disk.  It has been understood for some time that a concentration gradient can lead to fluid flow which will push a suspended particle (53); this effect is known as ``diffusiophoresis."  In the case of actin-polymerization-driven motion of a particle such as a bacterium, bead or disk, the particle itself gives rise to the non-equilibrium concentration gradient, so the phenomenon is an example of ``self-diffusiophoresis" (54).  A recent experiment observing motility of colloids coated on one side with platinum that catalyzes a chemical reaction in solution is an illustration of a very similar phenomenon (55).  

Now that we have identified a potential mechanism, we must ask whether it would be significant in the real system, which we will take to be a micron-sized bead moving in a cell extract.   The real system differs from our simulation in two very important ways.  First, the parameter range is very different; the real system has a much lower actin concentration and depolymerization rate constant.  Second, there is fluid flow in the real system but not in our simulation.  To see whether the proposed mechanism is relevant to the real system, we estimate the speed resulting from the mechanism for realistic conditions within a back-of-the-envelope calculation, and compare it to the observed speed.  If it is within an order of magnitude or so of the experimentally observed speed, our candidate is a reasonable one for the mechanism of motility.   

To estimate the speed, we must first estimate the concentration gradient of actin near the surface of the disk.  Because there is a short-ranged repulsion between monomers and the disk, the concentration at the disk is approximately zero.  The concentration of filaments immediately behind the disk, on the other hand, has been estimated by previous calculations~(28, 29) to be roughly 1mM.  The scale of the rise is on the order of the mesh size of the branched network, roughly 50 nm, so we take the concentration gradient to be $\grad c \approx $ 1mM/50 nm.  

We must now calculate the resulting pressure gradient.  A crude estimate is based on the ideal gas result, where $\grad p \approx kT \grad c$.  In the real system, this pressure gradient will lead to an equal and opposite pressure gradient acting on the fluid, which will lead to fluid flow in the actin comet tail that pushes the disk forwards (recall that we assume that the disk prefers to have water near it rather than actin).  The magnitude of this flow velocity is related by Darcy's law to the pressure gradient of the actin via the permeability, $k$:
\begin{equation}
v \approx -\frac{k}{\eta} \grad p
\end{equation}
where $\eta \approx 2.4$ cP is the viscosity of cell extract (56).  The permeability of the actin comet tail can be estimated from calculations for random fiber networks (57) to be $k \approx 10^{-5} \mu$m$^2$, but this is quite uncertain; it is only clear that it should be quite low.  Putting this all together, we obtain a fluid flow speed of $v \approx 1 \mu$m/s.   The speed of the bead should be comparable.  This speed is within an order of magnitude of the observed speed, which is excellent agreement despite the considerable uncertainty in the permeability and pressure estimates.  This encouraging result suggests that self-diffusiophoresis is a good candidate for the origin of motility in actin-polymerization-driven systems.

The proposed mechanism of motility is falsifiable by a relatively straightforward experiment.   According to our simulations, the key to motility lies in the concentration gradient of actin near the disk, which decreases as one approaches the disk from behind because the disk repels actin.  This depends on the density of actin at the maximum, $\rho_{\rm max}$, which occurs roughly 30 nm behind the surface in our simulations (see Fig.~\ref{stress}(b)), as well as the density at the surface, $\rho_{\rm surf}$.   In the real system, N-WASP or Act-A at the surface not only activates Arp2/3 but also binds F-actin, giving rise to an increase in $\rho_{\rm surf}$ and therefore perhaps decreasing the speed.  As the coverage of N-WASP or Act-A increases, both $\rho_{\rm max}$ and $\rho_{\rm surf}$ presumably increase, leaving the difference relatively unaffected.  This may be why the speed has been observed to be relatively insensitive to the coverage of Act-A~(7) or N-WASP, at least at high coverage (34).   To test our proposed mechanism, we therefore propose the following experiment.  Suppose one adds another protein to the surface, in addition to N-WASP, that binds F-actin but does not activate Arp2/3.  By increasing the coverage of this second protein at fixed coverage of N-WASP, one should be able to increase $\rho_{\rm surf}$ without affecting $\rho_{\rm max}$.  This would decrease the concentration gradient, so we would predict that it would slow down the particle, and possibly even reverse its direction of motion.

The minimal model we have presented here was designed to capture the most important features of the dendritic nucleation model, and we have tested it by reproducing results from experiments on purified proteins.  One feature of the experimental system is missing--this is the binding of filaments to the protein that activates Arp2/3 complex (N-WASp, ActA, etc.).  The next step is to incorporate specific binding of filaments to the surface.  However, we note that our success in reproducing known non-monotonic trends with various proteins is encouraging, and suggests that binding may not be essential to understanding {\it all} features of actin-based motility.  

Once we have incorporated binding, the next step will be to incorporate bundling or crosslinking proteins and to use a curved surface.  These extensions will allow us to study situations in which the biology has been perturbed, such as  ActA mutants that can hop (58), bundled systems that still move after Arp2/3 has been removed (59), and systems that move faster or slower when crosslinking proteins have been added (31).

In summary, we have conducted the first physically-consistent simulations of actin-polymerization-driven motility.  These simulations are also the first to include semiflexible filaments and to qualitatively reproduce experimentally-measured, non-monotonic trends with the various proteins involved.  Our results suggest a new picture for the mechanism of motility that is experimentally falsifiable.

\section*{Methods}

Here we describe the model, which we solve numerically using Brownian dynamics methods.

\subsection*{Interactions}

All actin monomers, whether free or bound in filaments, are modeled as spheres of size $\sigma \equiv 5$nm that repel each other with a soft repulsive potential
\begin{equation}
\label{rep}
\Phi_{\rm R} = \frac{1}{2} K \sum_{\{ij\}} (R_{ij}-R_0)^2, \quad R_{ij}<R_0=1\sigma \\
\end{equation}
with $K=100\kb T/\sigma^2$.
Monomers within filaments interact with each other via a bond potential 
\begin{equation}
\label{stretch}
\Phi_{\rm S} = \frac{1}{2}K\sum_{\{ij\}}(R_{ij}-D_0)^2, \quad R_{ij} > D_0=1\sigma \\
\end{equation}
with $K=100 \kb T/\sigma^2$ as in Eq.~\ref{rep}.  We introduce a bending potential that imparts stiffness to the filament (60):
\begin{equation}
\label{bend}
\Phi_{\rm B} = \frac{1}{2}K_B\sum_i (\cos(\theta_i)-\cos(\theta_0))^2,
\end{equation}
We use $K_B=100 \kb T$ in most of our runs, but have also explored the effect of filament stiffness by using $K_B=0\kb T$ and $K_B=1000\kb T$.  In Eq.~\ref{bend},  $\theta$ is the angle between the bond connecting monomer $i-1$ to monomer $i$ and the bond connecting monomer $i$ to monomer $i+1$ along a filament and $\theta_0= 0^\circ$ (see Fig.~\ref{fig1}(a)).  Note that $i=1$ corresponds to the pointed end.  If monomer $i$ is tagged by Arp2/3 complex and is at a y-junction (37, 61, 62), there is also a bending potential of the same form as Eq.~\ref{bend}, where $\theta$ is the angle between the bond connecting monomer $i$ at the junction to monomer $i+1$ on the branch, and the bond connecting monomer $i-1$ on the parent filament to monomer $i$ at the junction (Fig.~\ref{fig1}(b)).  In that case, $\theta_0=70^\circ$ (37).

For the moving surface, we use a flat disk of thickness $\sigma$ and radius $10 \sigma$ (63).  Monomers are repelled from the disk with a potential similar to Eq.~\ref{rep}.  Note that we have not included any attractive interaction between filaments and the disk.  As a result, the branched network is not attached to the disk, unlike the experimental system (29, 38-41).  
Model calculations (42) suggest that the speed $v(E_b)$ at binding energy $E_b$ is given by $v(E_b)=\alpha (E_b) v(E_b=0)$, where $\alpha(E_b)$ does not depend on $v(E_b=0)$.    This paper focuses on the physical origin of $v(E_b=0)$.  We note that even without including binding, we are able to reproduce nontrivial, qualitative trends observed experimentally.  Thus, it appears that binding may not be essential to understanding all aspects of motility.   Here, we have also neglected crosslinking of the filaments since it is known that neither crosslinking nor bundling proteins are needed for motility (8).

\subsection*{Branching}

In our model, the Arp2/3 complex is treated as a point particle that is generated (activated) at the center of one side of the disk and diffuses away from it.  We have also generated Arp2/3 at random points on one side of the disk and found that this makes no difference to the speed.  By generating Arp2/3 from one side of the disk but not the other, we break symmetry.  As a result, the branched actin network self-assembles on one side of the disk and drives it, on average, in a specific direction (which we define as the $+z$ direction).  If Arp2/3 collides with the disk it is reflected without exerting a force on the disk.  If Arp2/3 collides with a monomer in a filament, it sticks to it and activates the monomer for branching.  The Arp2/3 remains stuck to the branching monomer until the branch falls off, the branching monomer is depolymerized, or the Arp2/3 spontaneously dissociates.  Once it detaches from the monomer, it is regenerated near the disk.  This procedure is designed to generate a physically reasonable Arp2/3 distribution near the disk surface (34, 63) without imparting forces to it as an artifact.  We have confirmed this by running simulations with $K_+$ set to zero so that polymerization cannot occur.  In this case, the emission of Arp2/3 from the disk does not lead to any motion of the disk.

Note that we do not restrict branching to the
barbed end.  Because the number of barbed-end monomers is low compared to the total number of monomers in filaments, side branching 
(10, 37, 65)
is the dominant branching mechanism in our model. 

\subsection*{Equations of motion}
All particles in our system (free monomers, monomers in filaments, Arp2/3 and the disk) evolve 
according to Brownian Dynamics (Eq.~\ref{bd}) (66), with corresponding phenomenological
fricion constants $\zeta$ and stochastic random forces $F$ (Eq.~\ref{sf}).   Thus, all free monomers, filaments, Arp2/3 and the disk fluctuate in position due to the stochastic random forces acting on them.  In addition, they are subjected to forces due to their interactions with each other:
\begin{equation}
\zeta_i \frac{{\rm d} X_i}{{\rm d}t} = - \nabla_i (\Phi_{\rm R} + \Phi_{\rm
S}+\Phi_{\rm B}) + F_i
\label{bd}
\end{equation}
\begin{equation}
\left<F_i\right>=0, \quad
\left<F_i(t) F_j(t')\right> = 6k_{\rm B} T \zeta_i \delta(t-t')\delta_{ij}
\label{sf}
\end{equation}

The friction constant $\zeta_0$ of an actin monomer 
is taken to be $\zeta_0 = 3\pi \sigma \eta$, where $\eta$ is the viscosity of the medium.  The friction constant of the disk is taken to be $\zeta_D=20 \zeta_0$, and that for the Arp2/3 is taken to be $\zeta_0$. 

We convert our results to real units as follows.  The unit of length in our model is the size of the actin monomer, $\sigma = 5$nm.   We take the viscosity to be 2.4cP, as measured experimentally for cell extracts (56).  This yields a monomer diffusion coefficient of $D \equiv  \kb T/\zeta_0 = 36\mu$m$^2$/s and a characteristic time unit of $\tau \equiv \sigma^2/(2D) = 0.35\mu$s.

\subsection*{Boundary conditions}
We use periodic boundary conditions.  The disk is constrained to move in the $z$-direction only.  Most of our results are for a system of size 40$\sigma \times 40\sigma \times 80\sigma$.
Unless explicitly stated otherwise, we find no discernible differences for system sizes 80$\sigma \times 80 \sigma \times 80\sigma$ and
40$\sigma \times 40\sigma \times160 \sigma$ under standard conditions
(Table 1).

\subsection*{Biochemistry}
The next step is to include the self-assembly/biochemistry of the dendritic
nucleation model.
Our algorithm for actin polymerization is similar to that of Gelbart et al. for 
nano-colloids (67).   We allow polymerization only at the barbed end or at a branching monomer tagged by Arp2/3, and allow depolymerization only at the pointed end.  

\begin{figure}[tbh]
\centering{
\psfig{file=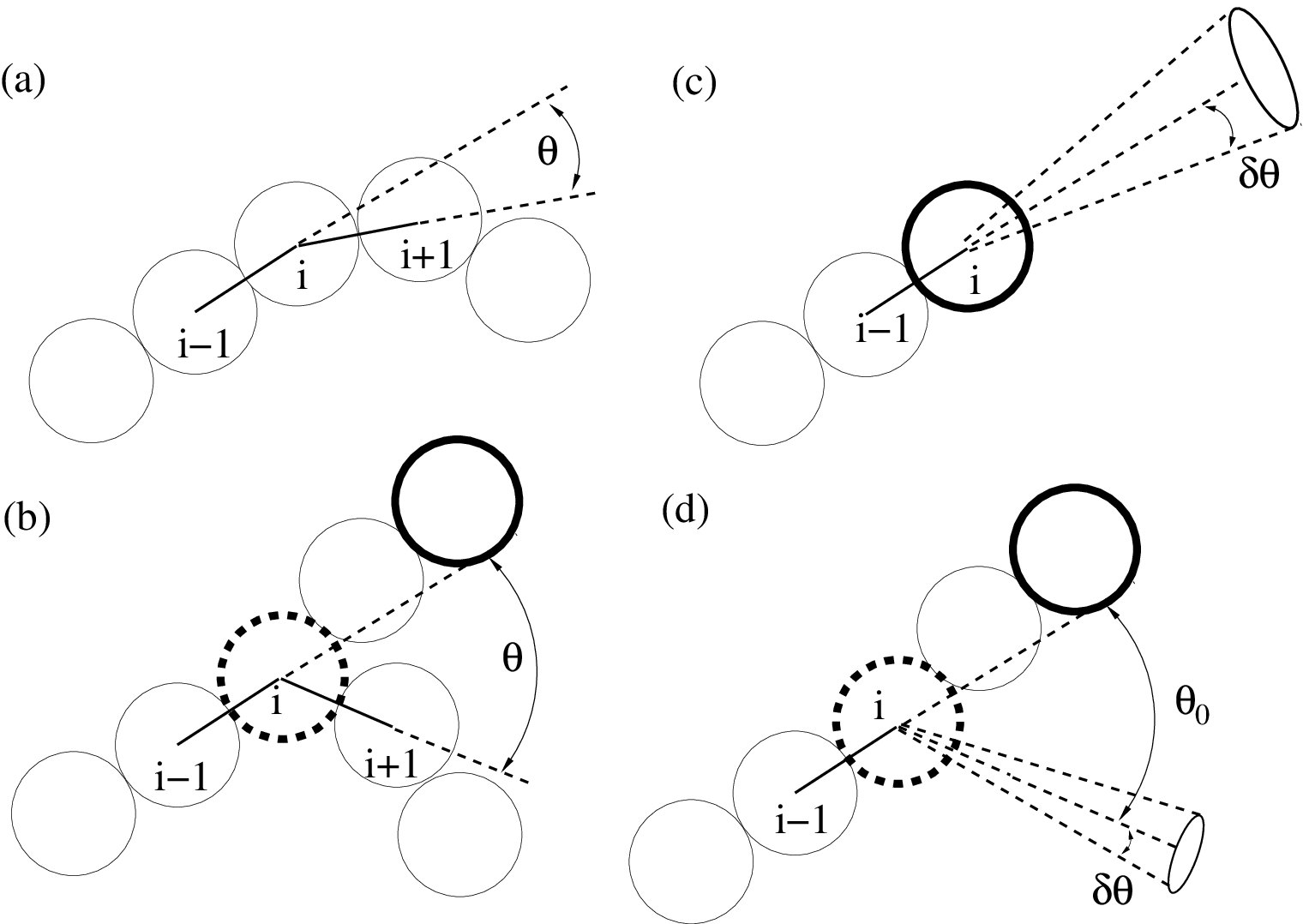,width=7.5cm,height=5cm}
}
\caption{
}
\label{fig1}
\end{figure}

Polymerization occurs when 
the center of a diffusing free monomer $j$ is within a
distance $R_{ij}$ of the monomer $i$ at the growing end of a filament (solid rimmed circle in Fig.~\ref{fig1}(c)), such that 
\begin{equation}
(\sigma - \delta r) < R_{ij} < \sigma.
\label{Rcapture}
\end{equation}
In addition to satisfying Eq.~\ref{Rcapture}, a free monomer $j$ must also lie 
within the angular cone 
\begin{equation}
|\cos(\theta)-\cos(\theta_0)| < \delta\theta
\label{thetacapture}
\end{equation}
of monomer $i$.  The angle $\theta$  is the angle between the vector 
from monomer $i-1$ preceeding monomer $i$ on the filament to monomer $i$ 
and the vector connecting monomer $i$ to free monomer $j$  (see Fig.~\ref{fig1}(c)).
Here, $\theta_0=0^\circ$. 

Polymerization also occurs when the center of free monomer $j$ is within $R_{ij}$ of monomer $i$ that has been tagged by Arp2/3 as a branching monomer (dashed rimmed circle in Fig.~\ref{fig1}(d)).  In that case, monomer $j$ must satisfy Eqs.~\ref{Rcapture} and ~\ref{thetacapture} with $\theta_0=70^\circ$ (see Fig.~\ref{fig1}(d)).  

In all cases, we choose $\delta r$ and $\delta \theta$ in Eqs.~\ref{Rcapture} and ~\ref{thetacapture}
such that the potential energy change due to polymerization is
small relative to the thermal energy $k_B T$.  Our 
choice of parameters ($\delta r = 0.1\sigma$,
$\delta \theta=0.02$) affects the effective polymerization rate but does not otherwise influence our results.  To verify this,
we have carried out a systematic calibration of the simulation
as a function of polymerization rate (Appendix).  

Depolymerization is simulated using a first order rate constant, $K_{-}$.
During each time step, $\Delta t=0.001 \tau$, {\it each} pointed end
is checked for depolymerization as follows:  a uniformly distributed random number between [0,1] is chosen and compared to the probability for dissociation during that time step, $K_{-}\Delta t$.
If the number is smaller 
than $K_{-}\Delta t$ the bond is broken to free the pointed-end monomer.  Capping is treated similarly; the probability for a barbed end to
become capped during a time step is $k_{\rm C+}\Delta t$, 
where $k_{\rm C+}$ is 
the pseudo-first order rate
constants for capping.  The probability for a barbed end to become uncapped 
is $K_{C-}\Delta t$,
where $K_{C-}$ is the first order rate for uncapping.  
Likewise, in each time step a branch can dissociate from its parent filament with
probability $K_{\rm d} \Delta t$ where $K_{\rm d}$ is the debranching rate. 
Finally, we note that we do not include ADF explicitly, but instead vary the depolymerization rate (68).    

\section*{Acknowledgments}
We thank T.~Haxton, T. C. Lubensky,  D.~J.~Pine, J.~M.~Schwarz and D.~Vernon for helpful discussions.  The support of the NSF through 
CHE-0613331 and the Penn MRSEC, DMR-0520020, is gratefully acknowledged. 

\section*{Appendix: Polymerization Parameter and Calibration}

It is well known that there is a change of free energy during polymerization; indeed, this is why polymerization occurs in the first place (4).  The system gains energy by polymerizing, and gains entropy by depolymerizing.  These free energy changes are directly related to the rate constants for polymerization and depolymerization.  In the steady state system with a moving surface, energy is continually added to the system because the system does not obey detailed balance; the system is driven out of equilibrium and the polymerization rate is much higher relative to the depolymerization rate than it would be in equilibrium.  In the real system,  ATP hydrolysis and Arp2/3 activation provide this additional energy. 

In our simulation, we do not explicitly include the free energy changes upon polymerization and depolymerization.  Rather, we introduce rate constants that implicitly depend on those free energy changes.  In our steady state system with a moving surface, the ratio of the polymerization rate to depolymerization rate has a constant value that exceeds the equilibrium constant, signifying that the system is out of equilibrium.

In addition to the free energy change upon polymerization, which we take into account using a rate constant, we could also introduce a change in the {\it mechanical} energy of a filament upon polymerization.  This could be done, for example, by storing energy in distortions of the filament.
Any mechanical energy added through polymerization would give
rise to forces that could lead to motility.   An important question is whether polymerization in itself, with no mechanical energy change in the filaments due to polymerization, can give rise to motility.  To address this question, we have designed our simulation model so that minimal mechanical energy is introduced into the filament upon polymerization.  In this appendix, we will show that it is not {\it necessary} to introduce a mechanical energy change in the filaments in order to obtain motility.  We note, however, that we cannot rule out the possibility that such a change occurs in the real system.

It has been suggested that ATP hydrolysis may occur during Arp2/3-mediated polymerization at the moving surface (17, 18).   In absence of direct experimental evidence of such a process, however, we prefer to concentrate on the simplest possible case, where no mechanical energy is added to the system even during Arp2/3-mediated polymerization, to see whether motility and reasonable force generation can still occur.

We cannot completely eliminate any addition of mechanical energy to filaments during the polymerization process.  However, we can minimize it as follows.  We have chosen the spring constant for monomer-monomer repulsion (Eq.~\ref{rep}) to be the same as the spring constant holding monomers together in filaments (Eq.~\ref{stretch}).  Thus, when a new bond is formed, the repulsive harmonic interaction is replaced by a full harmonic potential with no energy change at any value of the polymerization parameter $\delta r$ in Eq.~\ref{Rcapture}.
However, it is impossible to avoid a mechanical energy change in the filament due to bending of the filament (Eq.~\ref{bend}).  The amount of energy change is determined by the parameter $\delta \theta$ in Eq.~\ref{thetacapture}, and is nonzero as long as $\delta \theta \neq 0$.
We minimize the effect of Eq.~\ref{bend} by choosing 
a small value for $\delta \theta$.  To verify that the resulting small change of bending energy does not significantly affect the speed, we also carry out a 
systematic calibration, as follows.

The range $\delta \theta$ affects not only the change of bending energy stored in the filament due to polymerization, but also the polymerization rate itself.  Larger values of $\delta \theta$ lead to larger values of $K_+$.  Both the change in mechanical energy and the polymerization rate can, in principle, affect the speed.  Here we check whether the dominant contribution to the change in speed with changing $\delta \theta$ arises from the change in $K_+$, and not the change of bending energy.  To check this, we first calculate a "calibration curve" for speed as function of polymerization rate for
$\delta\theta=0.02$.  If changing $\delta\theta$ affects only the polymerization
rate, then for different values of $\delta\theta$, corresponding to different
polymerization rates, we should obtain speeds that lie
somewhere on the calibration curve. 

In order to calculate the calibration curve, we must first measure the polymerization rate.  To do this, we construct a system starting with a
fixed small concentration of dimers, free monomers, and no disk.  We turn off branching, capping and depolymerization and measure the rate of depletion of free monomers.  The free monomer concentration as function of time is a first order decay, so we fit it to $b_0\exp^{-K_{+}c_0 t}$ where $b_0$ is the
initial concentration of free monomers and $c_0$ is the initial concentration
of dimers.  The fitting parameter $K_{+}$ is the polymerization rate.
The value of $K_{+}$ for our standard setup is listed in Table 1. 

\begin{figure}[tbh]
\centering{
\psfig{file=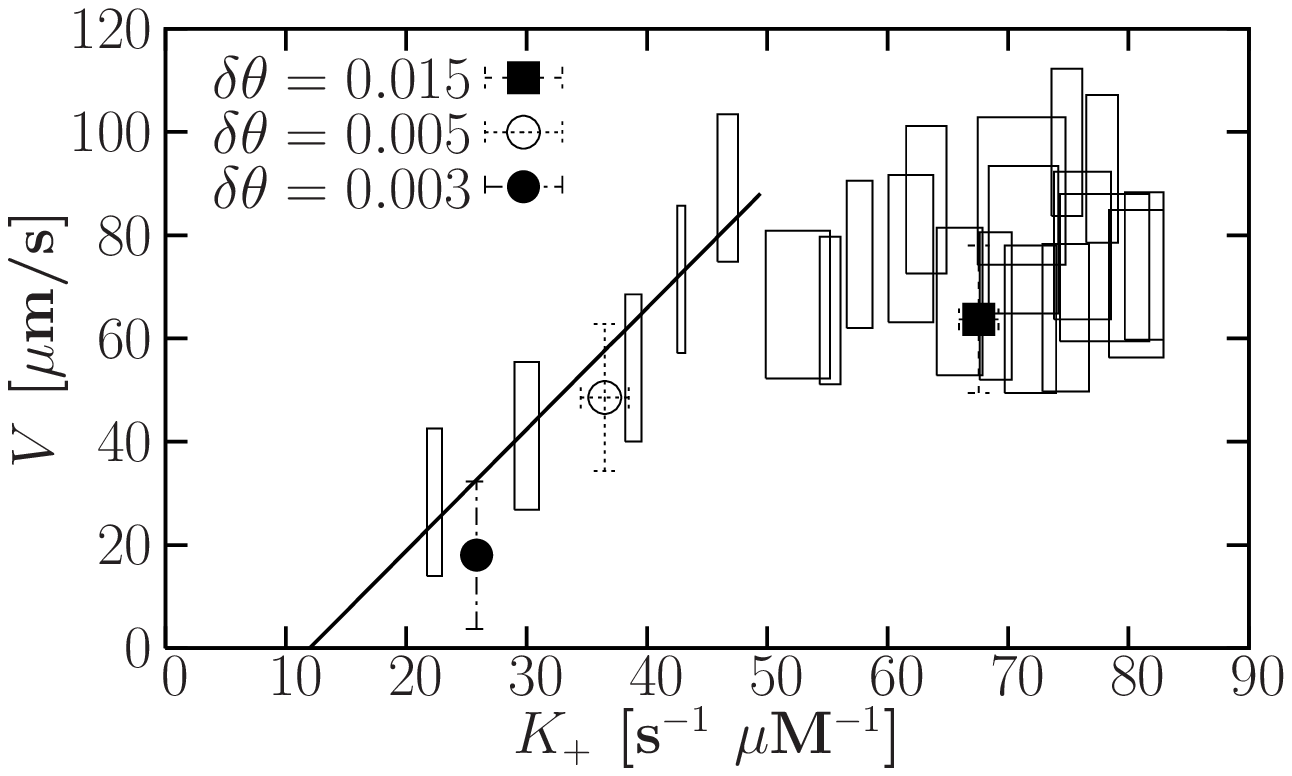,width=8.5cm}
}
\caption{
}
\label{cal}
\end{figure}

The next step in calculating the calibration curve is to vary the
polymerization rate without changing the value of $\delta \theta$.  This can be done without changing the energy of the system by introducing a probability $P_b \le 1$ for capture of a monomer by the barbed end, given that the free monomer satisfies the conditions of Eqs.~\ref{Rcapture}-\ref{thetacapture}.  In our standard runs, we use $P_b=1$, so we can only decrease the polymerization rate by using $P_b<1$.    The resulting curve for speed vs. polymerization rate is shown in Fig.~\ref{cal} (rectangular points).

With the calibration curve now in hand, we compute the polymerization rate and velocity
for three different values of $\delta \theta$.
As shown in Fig.~\ref{cal}, 
the speeds observed for $\delta \theta = 0.003, 0.005$, and $0.015$ fall on
the calibration curve, as expected.  This result demonstrates that $\delta \theta$ affects only the polymerization rate, and that the motility is not caused by sudden changes in the bending energy of filaments undergoing polymerization.  In other words, changing $\delta \theta$ only affects the speed through $K_+$ at small $\delta \theta$; there is no significant contribution from the change of bending energy stored in the filament.  Thus, we conclude that it is not necessary to include an explicit mechanical energy change upon polymerization in order to obtain motility.

Fig.~\ref{cal} shows that the speed of the disk increases linearly at low
polymerization rate and saturates at high polymerization rate.  
The saturation value of $K_+$ is related to [Arp2/3]; at low $K_+$, 
the velocity is limited by the rate of creation of new growing ends.
A straight line fit to the low-polymerization-rate 
portion of the curve shows that
a threshold polymerization rate is needed to obtain a nonzero speed. 
This threshold rate yields an estimate of the critical
actin concentration required for motility, 
given $K_-$ from Table 1.   We find that the critical
actin concentration is
$K_{-}/K_{+}\sim 2$mM, consistent with what we found before in
Fig.~4(d).

\clearpage

\section*{References}

1. Tilney, L. G., and D. A. Portnoy, 1989. Actin filaments and the growth,
movement, and spread of the intracellular bacterial parasite. J. Cell
Biol. 109:1597Ð1608.

2. Condeelis, J., 1993. Life at the leading edge: the formation of cell
protrusions. Annu. Rev. Cell Biol. 9:411Ð444.

3. Pollard, T. D., L. Blanchoin, and R. D. Mullins, 2000. Molecular mech-
anisms controlling actin filament dynamics in nonmuscle cell. Annu.
Rev. Biophys. Biomol. Struct. 29:545Ð576.

4. Hill, T. L., and M. W. Kirschner, 1982. Bioenergetics and kinetics of
microtubule and actin filament assembly-disassembly. Int. Rev. Cytol.
78:1Ð125.

5. Oosawa, F., and S. Asakura, 1975. Thermodynamics of the Polymeriza-
tion of Protein. Academic Press Inc.

6. Borisy, G. G., and T. M. Svitkina, 2000. Actin machinery: pushing the
envelope. Curr. Opin. Cell Biol. 12:104Ð112.

7. Cameron, L. A., M. J. Footer, A. van Oudenaarden, and J. A. Theriot,
1999. Motility of ActA protein-coated microspheres driven by actin
polymerization. Proc. Natl. Acad. Sci. USA 96:4908Ð4913.

8. Loisel, T. P., R. Boujemaa, D. Pantaloni, and M.-F. Carlier, 1999. Re-
constitution of actin-based motility of Listeria and Shigella using pure
proteins. Nature 401:613Ð616.

9. Bernheim-Groswasser, A., S. Wiesner, R. M. Golsteyn, M.-F. Carlier,
and C. Sykes, 2002. The dynamics of actin-based motility depend on
surface parameters. Nature 417:308Ð311.

10. Mullins, R. D., J. A. Heuser, and T. D. Pollard, 1998. The interaction
of Arp2/3 complex with actin: Nucleation, high affinity pointed end
capping, and formation of branching networks of filaments. Proc. Natl.
Acad. Sci. USA 98:6181Ð6186.

11. Cameron, L. A., P. A. Giardini, F. S. Soo, and J. A. Theriot, 2000.
Secrets of actin-based motility revealed by a bacterial pathogen. Nature
Rev. Mol. Cell Biol. 1:110Ð119.

12. Peskin, C. S., G. M. Odell, and G. F. Oster, 1993. Cellular motions and
thermal fluctuations: the Brownian ratchet. Biophys. J. 65:316Ð324.

13. Mogilner, A., and G. F. Oster, 1996. Cell motility driven by actin
polymerization. Biophys. J. 71:3030Ð3045.

14. Mogilner, A., and G. F. Oster, 2003. Force generation by actin poly-
merization II: The elastic ratchet and tethered filaments. Biophys. J.
84:1591Ð1605.

15. Burroughs, N. J., and D.Marenduzzo, 2005. Three-dimensional dynamic
Monte Carlo simulations of elastic actin-like ratchets. J. Chem. Phys.
123:174908.

16. Burroughs, N. J., and D. Marenduzzo, 2006. Growth of a semi-flexible
polymer close to a fluctuating obstacle: application to cytoskeletal actin
fibres and testing of ratchet models. J. Phys:Condens. Matt. 18:S357Ð
S374.

17. Dickinson, R. B., and D. L. Purich, 2002. Clamped-filament elongation
model for actin-based motors. Biophys. J. 82:605Ð617.

18. Dickinson, R. B., L. Caro, and D. L. Purich, 2004. Force generation by
cytoskeletal filament end-tracking proteins. Biophys. J. 87:2838Ð2854.

19. Dickinson, R. B., and D. L. Purich, 2006. Diffusion Rate Limitations in
Actin-Based Propulsion of Hard and Deformable Particles. Biophys. J.
91:1548Ð1563.

20. Carlsson, A. E., 2000. Force-velocity relation for growing biopolymers.
Phys. Rev. E 62:7082Ð7091.

21. Carlsson, J. Z. A. E., 2006. Growth of attached actin filaments. Eur.
Phys. J. E 21:209Ð222.

22. Gholami, A., J. Wilhelm, and E. Frey, 2006. Entropic forces generated
by grafted semiflexible polymers. Phys. Rev. E 74:041803.

23. Maly, I. V., and G. G. Borisy, 2001. Self-organization of a propulsive
actin network as an evolutionary process. Proc. Natl. Acad. Sci. USA
98:11324Ð11329.

24. Mogilner, A., and L. Edelstein-Keshet, 2002. Regulation of actin dy-
namics in rapidly moving cells: a quantitative analysis. Biophys. J.
83:1237Ð1258.

25. Schaus, T. E., E.W. Taylor, and G. G. Borisy, 2007. Self-organization of
actin filament orientation in the dendritic-nucleation/array-treadmilling
model. Proc. Natl. Acad. Sci. USA 104:7086Ð7091.

26. Carlsson, A. E., 2003. Growth velocities of branched actin networks.
Biophys. J. 84:2907Ð2918.

27. Gopinathan, A., K.-C. Lee, J. M. Schwarz, and A. J. Liu, 2007. Branch-
ing, capping, and severing in dynamic actin structures. Phys. Rev. Lett.
99:058103.

28. Carlsson, A. E., 2001. Growth of branched actin networks against ob-
stacles. Biophys. J. 81:1907Ð1923.

29. Gerbal, F., V. Laurent, A. Ott, M.-F. Carlier, P. Chaikin, and J. Prost,
2000. Measurement of the elasticity of the actin tail of Listeria monocytogenes
. Eur. Biophys. J. 29:134Ð140.

30. Gerbal, F., P. Chaikin, Y. Rabin, and J. Prost, 2000. An elastic analysis
of Listeria monocytogenes propulsion. Biophys. J. 79:2259Ð2275.

31. Paluch, E., J. van der Gucht, J.-F. Joanny, and C. Sykes, 2006. Defor-
mations in actin comets from rocketing beads. Biophys. J. 91:3113Ð3122.

32. Alberts, J. B., and G. M. Odell, 2004. In silico reconstitution of Listeria
propulsion exhibits nano-saltation. PLoS Biol. 2:2054Ð2066.

33. Burroughs, N. J., and D. Marenduzzo, 2007. Nonequilibrium-driven
motion in actin networks: comet tails and moving beads. Phys. Rev.
Lett. 98:238302.

34. Wiesner, S., E. Helfer, D. Didry, G. Ducouret, F. Lafuma, M.-F. Carlier,
and D. Pantaloni, 2003. A biomimetic motility assay provides insight
into the mechanism of actin-based motility. J. Cell Biol. 160:387Ð398.

35. Marchand, J.-B., P. Moreau, A. Paoletti, P. Cossart, M.-F. Carlier, and
D. Pantaloni, 1995. Actin-based movement of Listeria monocytogenes:
actin assembly results from the local maintenance of uncapped filament
barbed ends at the bacterium surface. J. Cell Biol. 130:331Ð343.

36. Lee, K.-C., and A. J. Liu, 2008. Numerical simulations of actin-
polymerization-driven motility. ACS Symposium To be published.

37. Blanchoin, L., K. J. Amann, H. N. Higgs, J.-B. Marchand, D. A. Kaiser,
and T. D. Pollard, 2000. Direct observation of dendritic actin filament
networks nucleated by Arp2/3 complex and WASP/Scar proteins. Nature
404:1007Ð1011.

38. Kuo, S. C., and J. L. McGrath, 2000. Steps and fluctuations of Listeria
monocytogenes during actin-based motility. Nature 407:1026Ð1029.

39. Upadhyaya, A., J. R. Chabot, A. Andreeva, A. Samadani, and A. van
Oudenaarden, 2003. Probing polymerization forces by using actin-
propelled lipid vesicles. Proc. Natl. Acad. Sci. USA 100:4521Ð4526.

40. Giardini, P. A., D. A. Fletcher, and J. A. Theriot, 2003. Compression
forces generated by actin comet tails on lipid vesicles. Proc. Natl. Acad.
Sci. USA 100:6493Ð6498.

41. Marcy, Y., J. Prost, M.-F. Carlier, and C. Sykes, 2004. Forces generated
during actin-based propulsion: A direct measurement by micromanipu-
lation. Proc. Natl. Acad. Sci. USA 101:5992Ð5997.

42. Gopinathan, A., and A. J. Liu In preparation.

43. Theriot, J. A., T. J. Mitchison, L. G. Tilney, and D. A. Portnoy, 1992.
The rate of actin-based motility of intracellular Listeria monocytogenes
equals the rate of actin polymerization. Nature 357:257Ð260.

44. Frey, E., K. Kroy, and J. Wilhelm, 1998. Physics of solutions and net-
works of semiflexible macromolecules and the control of cell function
ArXiv:cond-mat/9808022.

45. Gardel, M. L., J. H. Shin, F. C. MacKintosh, L. Mahadevan, P. Mat-
sudaira, and D. A. Weitz, 2004. Elastic behavior of cross-linked and
bundled actin networks. Science 304:1301Ð1305.

46. Carlsson, A. E., 2004. Structure of autocatalytically branched actin
solutions. Phys. Rev. Lett. 92:238102.

47. Blanchoin, L., T. D. Pollard, and R. D. Mullins, 2000. Interactions of
ADF/Cofilin, Arp2/3 complex, capping protein and profilin in modelling
of branched actin filament networks. Curr. Biol. 10:1273Ð1282.

48. Noireaux, V., M. Golsteyn, E. Friederich, J. Prost, C. Antony, D. Lou-
vard, and C. Sykes, 2000. Growing an actin gel on spherical surfaces.
Biophys. J. 78:1643Ð1654.

49. Plastino, J., I. Lelidis, J. Prost, and C. Sykes, 2004. The effect of
diffusion, depolymerization and nucleation promoting factors on actin
gel growth. Eur. Biophys. J. 33:310Ð320.

50. Roberts, T. M., and M. Stewart, 2000. Acting like actin: the dynamics
of the nematode major sperm protein (MSP) cytoskeleton indicate a
push-pull mechanism for amoeboid cell motility. J. Cell Biol. 149:7Ð12.

51. Bottino, D., A. Mogilner, T. Roberts, M. Stewart, and G. Oster, 2002.
How nematode sperm crawl. J. Cell Biol. 115:367Ð384.

52. King, K. L., M. Stewart, and T. M. Roberts, 1994. Supramolecular
assemblies of the Ascaris suum major sperm protein (MSP) associated
with amoeboid cell motility. J. Cell Sci. 107:2941Ð2949.

53. Anderson, J. L., 1989. Colloid transport by interfacial forces. Annu.
Rev. Fluid Mech. 21:61Ð99.

54. Golestanian, R., T. B. Liverpool, and A. Ajdari, 2005. Propulsion of
a molecular machine by asymmetric distribution of reaction products.
Phys. Rev. Lett. 94:220801.

55. Howse, J. R., R. A. L. Jones, A. J. Ryan, T. Gough, R. Vafabakhsh,
and R. Golestanian, 2007. Self-motile colloidal particles: from directed
propulsion to random walk. Phys. Rev. Lett. 99:048102.

56. Cameron, L. A., J. R. Robbins, M. J. Footer, and J. A. Theriot, 2004.
Biophysical parameters influence actin-based movement, trajectory, and
initiation in a cell-free System. Mol. Biol. Cell 15:2312Ð2323.

57. Koponen, A., D. Kandhai, E. Hellen, M. Alava, A. Hoekstra, M. Kataja,
K. Niskanen, P. Sloot, and J. Timonen, 1998. Permeability of three-
dimensional random fiber webs. Phys. Rev. Lett. 80:716Ð719.

58. Lasa, I., E. Gouin, M. Goethals, K. Vancompernolle, V. David, J. Van-
dekerchhove, and P. Cossart, 1997. Identification of two regions in the
N-terminal domain of ActA involved in the actin comet tail formation
by Listeria monocytogenes. EMBO 16:1531Ð1540.

59. Brieher, W. M., M. Coughlin, and T. J. Mitchison, 2004. Fascin-
mediated propulsion of Listeria monocytogenes independent of frequent
nucleation by the Arp2/3 complex. J. Cell Biol. 165:233Ð242.

60. Rapaport, D. C., 1995. The Art of Molecular Dynamics Simulations.
Cambridge University Press.

61. Svitkina, T. M., and G. G. Borisy, 1999. Arp2/3 complex and actin
depolymerizing factor/Cofilin in dendritic organization and treadmilling
of actin filament array in lamellipodia. J. Cell Biol. 145:1009Ð1026.

62. Cameron, L. A., T. M. Svitkina, D. Vignjevic, J. A. Theriot, and G. G.
Borisy, 2001. Dendritic organization of actin comet tails. Curr. Biol.
11:130Ð135.

63. Schwartz, I. M., M. Ehrenberg, M. Bindshcadler, and J. M. McGrath,
2004. The role of substrate curvature in actin-based pushing forces.
Curr. Biol. 14:1094Ð1098.

64. Bailly, M., F. Macaluso, M. Cammer, A. Chan, J. E. Segall, and J. S.
Condeelis, 1999. Relationship between Arp2/3 Complex and the barbed
ends of actin filaments at the leading edge of carcinoma cells after epi-
dermal growth factor stimulation. J. Cell Biol. 145:331Ð345.

65. Carlsson, A. E., M. A. Wear, and J. A. Cooper, 2004. End versus side
branching by Arp2/3 complex. Biophys. J. 86:1074Ð1081.

66. Allen, M. P., and D. J. Tildesley, 1987. Computer Simulation of Liquids.
Oxfor University Press.

67. Gelbart, W. M., R. P. Sear, J. R. Heath, and S. Caney, 1999. Array
formation in nano-colloids: Theory and experiment in 2D. Faraday
Discuss. 112:299Ð307.

68. Carlier, M.-F., V. Laurent, J. Santolini, R. Melki, D. Didry, G.-X. Xia,
Y. Hong, N.-H. Chua, and D. Pantaloni, 1997. Actin depolymerizing fac-
tor (ADF/Cofilin) enhances the rate of filament turnover: Implication
in actin-based motility. J. Cell Biol. 136:1307Ð1322.

69. Podolski, J. L., and T. L. Steck, 1990. Length distribution of F-actin in
Dictyostelium discoideum. J. Biol. Chem. 265:1312Ð1318.

\clearpage
\section*{Tables}
\begin{table}[tbh]
\centering{
\begin{tabular}{rrrr}
Parameter &{\it in vitro} Exp. (ref)&Simulated &\\
\hline \hline 
$l_p$ &0.5-15 $\mu$ m (13) & 0.5 $\mu$m&\\
\hline
$l_{\rm ave}$& 0.1-1 $\mu$m (64, 69) & 0.1 $\mu$m &\\
\hline 
typical bead diameter& 0.2-2 $\mu$m (56) & 0.1 $\mu$m&\\
\hline
viscosity $(\eta)$ & 2.4 cP (56) & 2.4 cP &\\
\hline
$D=\kb T /3\pi\eta\sigma$ & 36 $\mu$m$^2$/s & 36 $\mu$m$^2$/s&\\
\hline
$K_+$ & $$11.6 $\mu$M$^{-1}$s$^{-1}$ (3) &  63 $\mu$M$^{-1}$s$^{-1}$  &\\ 
\hline
$K_-$& 0.3 s${^-1}$ (3) &  28600 s$^{-1}$ &\\
\hline
[G-Actin]& 7 $\mu$M (8) & 5000$\mu$M &\\
\hline
${\frac{K_+[{\rm G-Actin}]}{K_-}}$& 270 & 11 &\\
\hline
$K_{a}$& ?$\mu$M$^{-1}$s$^{-1}$  &  $\sim K_{+}$  &\\
\hline
$K_{d}$& 0.002 s$^{-1}$ (47) & 28600 s$^{-1}$ &\\
\hline
[Arp2/3]& 0.1 $\mu$M (8) & 17 $\mu$M &\\
\hline
$\frac{K_a[{\rm Arp2/3}]}{K_d}$ (65) & N/A & 0.037 &\\
\hline
$K_{C+}$& 8 $\mu$M$^{-1}$s$^{-1}$  (65) & \bf ---  &\\
\hline
$K_{C-}$& 0.00042 s$^{-1}$  (65)  & 0 s$^{-1}$ &\\
\hline
[Cap]& 0.1 $\mu$M (8) & \bf ---  &\\
\hline
$k_{\rm C+} = K_{C+}$[Cap]& 0.8 s${^-1}$ & 14300 s$^{-1}$ &\\
\hline \hline 
&&&
\end{tabular}

\caption{
Values of the parameters used in the simulations compared to those in
experiments.
}
}
\label{tab1}
\end{table}

\clearpage
\section*{Figures}

\subsubsection*{Figure 1}
(a) (Color online) Disk displacement as function of time for three values of the bending stiffness of filaments, $K_B$=1000 (dotted), 100 (solid) and 0 (dashed).   The vertical dashed lines show the times corresponding to snapshots I-III, namely 70 $\mu$s, 700 $\mu$s and 2100 $\mu$s, respectively, after the simulation started with monomers and 5\% dimers distributed randomly.  Snapshots:  Green spheres are monomers in filaments, red spheres are monomers tagged by Arp2/3 for branching, and blue spheres are the pointed-ends of filaments.  G-actin monomers are not
shown.  The disk is purple.
The black box in each frame marks
the boundary of the periodic box.  

\par
\par

\subsubsection*{Figure 2}
(a) Average force density on the monomers as a function of position in the frame of the moving disk, which is at $z=0$, denoted by the vertical dashed line.  Positive (negative) force 
implies monomers are being pushed to the right (left).   (b) Total local monomer density
$\rho(z)$ (solid), local filament monomer density (dashed) and local free monomer density (dotted).  

\par
\par

\subsubsection*{Figure 3}
Concentration dependence of speed.  (a) [Arp2/3] dependence.  Solid symbols correspond to runs done at the standard rates shown in Table 1.  Open symbols correspond to runs done with $K_-$, $K_d$ and [G-actin] reduced by a factor of 10 and $k_{C+}$ reduced by 5.  In both cases, there is clear non-monotonic behavior.  (b) Capping
rate dependence.  (c) Depolymerization rate dependence.  Symbols correspond to the cases in which Arp2/3 protects (solid) or does not protect (open) the pointed end from depolymerization.  (d) [G-actin]
dependence.  

\par
\par

\subsubsection*{Figure 4}
Schematic showing the definition of angles used in the bending potential energy in Eq.~3.   
(a)  For a monomer $i$ along the filament, there is a bending cost associated 
with changes of the angle $\theta$ away from $\theta_0=0$.   
(b)  If monomer $i$ is tagged by Arp2/3 and is at a y-junction, there 
is also a bending cost associated with changes of the angle $\theta$ with respect 
to $\theta_0=70^\circ$.  
(c)  A free monomer $j$ (not shown) can be added to monomer $i$ at a growing end if 
its center is within a range $\delta r$ of separations $R_{ij}$ 
such that $\sigma-\delta r<R_{ij}<\sigma$, and a range $\delta \theta$ of 
angles $\theta_{ij}$ around $\theta_0=0$ 
such that $|\cos \theta_{ij}-\cos\theta_0|<\delta\theta$.  
(d) A free monomer $j$ (not shown) can be added as the 
first monomer along a branch if monomer $i$ has been tagged by Arp2/3 complex, 
the separation $R_{ij}$ satisfies $\sigma-\delta r<R_{ij}<\sigma$, and 
the angle $\theta_{ij}$ satisfies $|\cos \theta_{ij}-\cos\theta_0|<\delta\theta$, 
where $\theta_0=70^\circ$.

\par
\par

\subsubsection*{Figure 5}
Velocity as function of polymerization rate calibration.  The data for the
calibration points (open rectangles) 
are obtained with $\delta \theta = 0.02$.  The size of each rectangle corresponds to the error associated with it.

\end{document}